\pdfoutput=1
\documentclass[pra,aps,twocolumn,floatfix,longbibliography,superscriptaddress]{revtex4-2}

\usepackage[usenames,dvipsnames]{xcolor}
\usepackage{amsmath}
\usepackage{amsfonts}
\usepackage{txfonts}
\usepackage{amssymb}
\definecolor{myurlcolor}{rgb}{0,0,0.7}
\definecolor{myrefcolor}{rgb}{0.1,0,0.9}
\usepackage{colortbl}
\newcommand{\dsl}[0]{\llbracket}
\newcommand{\dsr}[0]{\rrbracket}

\usepackage[
	breaklinks,
	colorlinks=true, 
	linkcolor=myrefcolor,
	citecolor=myurlcolor,
	urlcolor=myurlcolor
]{hyperref}
\usepackage{graphicx}
\usepackage{bbold}
\usepackage[makeroom]{cancel}        %
\usepackage{multirow}                %
\usepackage[normalem]{ulem}         %
\usepackage{array}
\usepackage{adjustbox}
\usepackage{mathtools}
\graphicspath{{./img/}}

\newcommand{\idg}[1]{{\bfseries #1)}}

\newcommand{\subfigimg}[3][,]{%
	\setbox1=\hbox{\includegraphics[#1]{#3}}%
	\leavevmode\rlap{\usebox1}%
	\rlap{\hspace*{2pt}\raisebox{\dimexpr\ht1-0.5\baselineskip}{{\bfseries \large\textsf{#2}}}}%
	\phantom{\usebox1}%
}

\definecolor{THc}{rgb}{0.9,0.3,0.2}

\begin{document}

\title{MaxSAT decoders for arbitrary CSS codes}

\author{Mohammadreza Noormandipour}
\email[Corresponding author: ]{mrn31@cam.ac.uk}
\affiliation{TCM Group, Cavendish Laboratory, Department of Physics, J J Thomson Avenue, Cambridge CB3 0HE, United Kingdom}
\affiliation{Artificial Intelligence Research Lab, Nokia Bell Labs; Broers Building, 21 J.J. Thomson Avenue, Cambridge, CB3 0FA, UK}

\author{Tobias Haug}
\email{tobias.haug@u.nus.edu}
\affiliation{Quantum Research Center, Technology Innovation Institute, Abu Dhabi, UAE}

\date{\today}

\begin{abstract}
Quantum error correction (QEC) is essential for operating quantum computers in the presence of noise. Here, we accurately decode arbitrary Calderbank-Shor-Steane (CSS) codes via the maximum satisfiability (MaxSAT) problem.
We show how to map quantum maximum likelihood problem of CSS codes of arbitrary geometry and parity check weight into MaxSAT problems. 
We incorporate the syndrome measurements as hard clauses, while qubit and measurement error probabilities, including biased and non-uniform, are encoded as soft MaxSAT clauses. 
For the code capacity of color codes on a hexagonal lattice, our decoder has a higher threshold and superior scaling in noise suppression compared to belief propagation with ordered statistics post-processing (BP-OSD), while showing similar scaling in computational cost.
Further, we decode surface codes and recently proposed bivariate quantum low-density parity check (QLDPC) codes where we find lower error rates than BP-OSD.
Finally, we connect the complexity of MaxSAT decoding to a computational phase transition controlled by the clause density of the MaxSAT problem, where we show that our mapping is always in the computationally ''easy`` phase. 
Our MaxSAT decoder can be further parallelised and implemented on ASICs and FPGAs, promising further substantial speedups. 
Our work provides a flexible platform towards practical applications on quantum computers.
\end{abstract}

\maketitle

\section{Introduction}
Quantum computers promise to solve important problems that cannot be addressed by classical computers~\cite{shor1999polynomial,nielsen2001quantum,point_set}. 
To run quantum computers even in the presence of noise, quantum error correction (QEC) codes suppress errors by encoding logical quantum information in many redundant physical qubits~\cite{shor1996fault,Mackay_1997,Resch_2021,girvin2023introduction}. 
An important class of QEC codes are Calderbank-Shor-Steane (CSS) codes~\cite{steane1996multiple,calderbank1996good}, which includes surface codes~\cite{bravyi1998quantum}, color codes~\cite{bombin2006topological} and bicycle quantum low-density parity check (QLDPC) codes~\cite{kovalev2013quantum,Bravyi_2024,voss2024trivariate}. Such CSS codes have been realized experimentally on various platforms~\cite{acharya2024quantum,google2023suppressing,bluvstein2024logical}.

In QEC, the key idea is to perform measurements on the physical system affected by errors, while not destroying the encoded logical information. From such syndrome measurements, one can infer whether an error has occurred, and which operation is best suited to correct the error to recover the original logical information. 
Here, a key challenge in QEC is decoding, i.e. inferring which is the most likely error that has occurred given the syndrome measurements.
Decoding is a challenging computational problem~\cite{iyer2015hardness} which has to be solved accurately in a as short time as possible. Various types of decoders have been proposed over the years, where decoders usually have to balance a tradeoff between high speed and high correction accuracy. 
This includes decoders based on tensor-networks~\cite{ferris2014tensor,chubb2021statistical,kaufmann2024blockbp}, neural networks~\cite{torlai2017neural}, annealing~\cite{fujisaki2022practical,takeuchi2023comparative,yugo2024ising}, belief-propagation~\cite{roffe2020decoding,panteleev2021degenerate}, perfect weight-matching~\cite{dennis2002topological,fowler2013minimum,higgott2023sparse}, renormalization group~\cite{duclos2010fast}, and union-find~\cite{delfosse2021almost}.

A recent approach proposed to decode via mapping to the maximum satisfiability (MaxSAT) problem~\cite{berent2023decoding}, which has been shown to provide a highly accurate decoder. However, so far the MaxSAT decoder has been limited to color codes and uniform noise models.

Here, we decode arbitrary CSS codes with high accuracy by mapping them into MaxSAT problems. 
Our decoder corresponds to the quantum maximum likelihood decoder~\cite{iyer2015hardness} with a completely general construction that works for any CSS code, including biased non-uniform noise models and noisy syndrome measurements.
We find that our decoder has the similar computational scaling as belief-propagation with ordered statistic post-processing (BP-OSD), while achieving superior decoding accuracy. 
In particular, for the color code our MaxSAT decoder has a higher threshold and better scaling of noise suppression with distance compared to BP-OSD, while showing similar scaling in computational time.
We also demonstrate higher accuracy for the recently proposed bicycle QLDPC codes by IBM~\cite{Bravyi_2024} and surface codes.
We relate the complexity of the MaxSAT decoding problem to a computational phase transition, showing that our mapping is in the easy computational phase.
Our decoder can be directly implemented on ASICs and FPGAs, which promises to dramatically speed up decoding and opens up the possibility of real-time decoding. 
Our approach provides a powerful and flexible framework to enhance QEC.

First, in Sec.~\ref{sec:Preliminaries} we introduce QEC, decoding and MaxSAT. Then, in Sec.~\ref{sec:MaxSAT_formulation} we show how to map the quantum decoding problem of general CSS codes into a MaxSAT problem and extend our construction to include noisy measurements. In Sec.~\ref{sec:clausedens}, we show the relationship between the decoding problem and the complexity phase transition of MaxSAT. We introduce our methods in Sec.~\ref{sec:methods} and numerical decoding results in Sec.~\ref{sec:result}. Finally, we discuss our results and conclude in Sec.~\ref{sec:discussion}.

\section{Preliminaries}\label{sec:Preliminaries}
We start by introducing the basics of QEC and MaxSAT.

\subsection{Quantum Maximum likelihood decoding}\label{sec:qmld}

A quantum error-correcting code $\dsl n, k, d\dsr$ has $n$ data qubits, $k$ logical qubits and code distance $d$. For a CSS code, the full parity check matrix 
\begin{equation}
H=\begin{pmatrix}
0 & H_X  \\
H_Z & 0
\end{pmatrix}
\end{equation}
is composed of parity check matrix $H_Z$ with only $Z$ Pauli parity checks, and $H_X$ with only $X$ checks. These stabilizer generators commute, which is ensured by the condition $H_X \cdot H_Z^\text{T} =0$. $H_X$ reveals phase ($Z$) errors, while $H_Z$ is responsible to detect bit-flip ($X$) errors. 
The parity checks act on $n$ data qubits, where we assume that the $j$th data qubit is subject to $X$ Pauli errors with probability $p_j^X$, $Z$ Pauli errors with probability $p_j^Z$ and $Y$ Pauli errors with probability $p_j^Y$.

We decode $H_X$ and $H_Z$ separately, where we now illustrate the decoding for the $m\times n$ matrix $H_Z$ ($H_X$ follows analogously)
\[
H_Z = \begin{pmatrix}
h_{11} & h_{12} & \cdots & h_{1n} \\
h_{21} & h_{22} & \cdots & h_{2n} \\
\vdots & \vdots & \ddots & \vdots \\
h_{m1} & h_{m2} & \cdots & h_{mn}
\end{pmatrix}
\]
where each element $h_{ij}$ is either $1$ or $0$, indicating whether qubit $j$ participates in parity check $i$ or not, and $m$ is the number of parity checks (rows), each having a Hamming weight $\omega_i=\sum_{j}h_{ij}$.  Physically, the $i$th parity check corresponds to measuring the Pauli string $\sigma_i=\otimes_{j=1}^n Z^{h_{ij}}$.
Measuring the $m$ parity checks yields a syndrome outcome $\boldsymbol{s} = (s_1, s_2, \cdots, s_m)$ with $s_i\in\{0,1\}$. Given the syndrome, the parity check constraint 
$\boldsymbol{s}= H_Z \boldsymbol{e}$, fixes the error configuration $\boldsymbol{e}=\{e_1,\dots, e_n\}$ with $e_j\in\{0,1\}$ indicating whether a bit-flip error has occurred for qubit $j$. 
Given the measured syndrome $\boldsymbol{s}$, one wants now  to figure out which is the actual error that has occurred. 
The quantum maximum likelihood decoding problem finds the error configuration with the highest probability given the constraints of the syndrome~\cite{iyer2015hardness}.

The probability that a bit-flip error configuration $\boldsymbol{e}$ has occurred is given by~\cite{higgott2023sparse}
\begin{equation}
    P(\boldsymbol{e})=\prod_{j=1}^n (1-p_j)^{1-\boldsymbol{e}_j} p_j^{\boldsymbol{e}_j}=\left[\prod_{j=1}^n (1-p_j)\right]\prod_{j=1}^n \left(\frac{p_j}{1-p_j}\right)^{\boldsymbol{e}_j}\,,
\end{equation}
where $p_j=p_j^X+p_j^Y$ is the probability that an $X$ or $Y$ error has affected the $j$th qubit, which both induce a bit-flip.
To find the most probable error $\boldsymbol{e}^{\text{ml}}$ given the measured syndrome $\boldsymbol{s}$, one has to maximize the log-likelihood
\begin{align}
    \boldsymbol{e}^{\text{ml}} = \arg \max_{\boldsymbol{e}} \{\ln(P(\boldsymbol{e})) \} &= \arg \max_{\boldsymbol{e}} \left(C-\sum_j w_j \boldsymbol{e}_j \right) \quad \label{eq:soft}\\
    \text{s.t. }\,  \boldsymbol{s}&=H_Z\boldsymbol{e} \,,\label{eq:hard}
\end{align} 
where we have the irrelevant constant $C=\sum_j \ln(1-p_j)$ and weights $w_j=\ln((1-p_j)/p_j)$ which represent the log-likelihood ratio between the probabilities of no error and a bit-flip error on the $j$th  qubit based on the prior $p_j$.

\subsection{MaxSAT}\label{sec:MaxSAT}

MaxSAT is an optimization variant of the Boolean satisfiability (SAT) problem where one aims to find an assignment of binary variables such that it satisfies maximum number of clauses.
The MaxSAT instance is constructed in a standard and widely used format called conjunctive normal form (CNF). For a summary of terms and conventions see Appendix~\ref{appen:A}. A \textit{k}-SAT instance, is a specific case of a Boolean satisfiability problem in which one is given a Boolean expression written in CNF form where each clause is constrained to $k$ literals. 
As an example, below we have a CNF formula for two boolean variables $e_1$ and $e_2$ (also called literals) consisting of three $2$-SAT clauses
\begin{equation}
    (e_1\lor e_2)\land(e_1\lor\lnot e_2)\land(\lnot e_1\lor e_2)\,,
\end{equation}
where $\lnot$ is the negation of the literal, $\lor$ indicates a logical OR and $\land$ a logical AND. In above example, the choice $e_1=e_2=1$ satisfies all three clauses, thus being the solution of the SAT problem.

\section{MaxSAT formulation of the decoding problem}\label{sec:MaxSAT_formulation}

We now map the decoding problem of arbitrary CSS code into a MaxSAT  problem.

\subsection{CNF Construction}\label{sec:CNF_construction}
As first step, we deal with the constraints due to the given syndrome~\eqref{eq:hard}, which turns out to be a SAT problem.
We expand~\eqref{eq:hard} to get $m$ binary addition (\textit{XOR}) equations
\begin{equation}\label{eq:constraint_i}
   s_i = h_{i1}e_1 \oplus h_{i2}e_2 \oplus \cdots \oplus h_{in}e_n \,.
\end{equation}
As variables $e_j$ with $h_{ij}=0$ do not contribute, each constraint $s_i$ involves only $\omega_i=\sum_j h_{ij}$ variables. 
The strategy to transform the above constraints to CNF clauses starts with giving a Boolean variable interpretation to the binary elements $e_j$, agreeing that True evaluates to binary value $1$.
In order to be able to construct $k$-SAT clauses, we need to break down the RHS of the \eqref{eq:constraint_i} to expressions involving less number of literals. 
In particular, we break down the $i^{\text{th}}$ constraint in~\eqref{eq:constraint_i} to smaller expressions by introducing a set of auxiliary literals $\{a_1, a_2, \cdots, a_{n-1}\}$ to save the result of the intermediate expressions in
\begin{equation}\label{eq:xor_breakdown}
    \left\{
    \begin{aligned}
    &a_1 = h_{i1} e_1 \oplus h_{i2} e_2 \\
    &a_2 = a_1 \oplus h_{i3} e_3 \\
    &a_3 = a_2 \oplus h_{i4} e_4 \\
    &\vdots \\
    &a_{n-3} = a_{n-4} \oplus h_{i(n-2)} e_{n-2} \\
    &s_i = a_{n-2} \oplus h_{in} e_n
    \end{aligned}
    \right.
\end{equation}
As mentioned before, the weight of each parity check is $\omega_i=\sum_j h_{ij}$, which means that out of $n$ elements $h_{i1}, h_{i2}, \cdots, h_{in}$ only $\omega_i$ of them are non-zero and need to be taken into account for evaluation of the $i^{\text{th}}$ constraint. The number of equations in~\eqref{eq:xor_breakdown} also would reduce to $\omega_i - 2$. %
Note that $h_{ij}$ and $s_i$ are known binary values in the above set of equations. There are various types of equations above that have a different form of translation to 3-SAT clauses. Tab.~\ref{tab:prescription} summarises the prescription for each type of equation to be translated to a 3-SAT CNF. We will later explain the motivation behind choosing clauses with length $3$.

\begin{table*}[!htbp]
\centering
\begin{adjustbox}{width=0.8\textwidth}
\begin{tabular}{||c|c|c||}
\hline
\hline
\textbf{Type of Equation} & \textbf{Example} & \textbf{Translation to 3-SAT CNF} \\
\hline
\cellcolor{blue!35} $n \geq 2: $ & \cellcolor{blue!35} & \cellcolor{blue!35}\\
\hline
\multirow{4}{*}{1. Equations involving 3 literals} & \multirow{4}{*}{$a_1 = e_1 \oplus e_2$} & $(\neg e_1 \lor \neg e_2 \lor \neg a_1) \land$ \\
 &  & $(e_1 \lor e_2 \lor \neg a_1) \land$ \\
 &  & $(\neg e_1 \lor e_2 \lor a_1) \land$ \\
 &  & $(e_1 \lor \neg e_2 \lor a_1)$ \\
\hline
\multirow{5}{*}{\shortstack{2. Last equation with $s_i = 0$ \\ Note: In this case we need the last auxiliary \\ variable $a_{n-1}$ to achieve the 3-SAT clause form}} & \multirow{5}{*}{$0 = a_{n-2} \oplus e_n$} & $(\neg a_{n-2} \lor e_n \lor a_{n-1}) \land$ \\
 &  & $(\neg a_{n-2} \lor e_n \lor \neg a_{n-1}) \land$ \\
 &  & $(a_{n-2} \lor \neg e_n \lor a_{n-1}) \land$ \\
 &  & $(a_{n-2} \lor \neg e_n \lor \neg a_{n-1})$ \\
 &  & \textit{Note: Logic equivalent to $(a_{n-2} = e_n)$} \\
\hline
\multirow{5}{*}{\shortstack{3. Last equation with $s_i = 1$ \\ Note: In this case we need the last auxiliary \\ variable $a_{n-1}$ to achieve the 3-SAT clause form}} & \multirow{5}{*}{$1 = a_{n-2} \oplus e_n$} & $(a_{n-2} \lor e_n \lor a_{n-1}) \land$ \\
 &  & $(a_{n-2} \lor e_n \lor \neg a_{n-1}) \land$ \\
 &  & $(\neg a_{n-2} \lor \neg e_n \lor a_{n-1}) \land$ \\
 &  & $(\neg a_{n-2} \lor \neg e_n \lor \neg a_{n-1})$ \\
 &  & \textit{Note: Logic equivalent to $(a_{n-2} \neq e_n)$} \\
\hline
\hline
\textbf{Weighted Soft Clauses} & \textbf{Example} & \textbf{Translation to 3-SAT CNF} \\
\hline
\multirow{4}{*}{\shortstack{1. If $w_j < 0$: literal \\ Note: To achieve a 3-SAT form we need 2 \\ auxiliary variables. Let's call them $b_{2j}$ and $b_{2j+1}$}} & \multirow{4}{*}{$|w_j|~e_j$} & $|w_j|~(e_j \lor b_{2j} \lor b_{2j+1}) \land$ \\
 &  & $|w_j|~(e_j \lor \neg b_{2j} \lor b_{2j+1}) \land$ \\
 &  & $|w_j|~(e_j \lor b_{2j} \lor \neg b_{2j+1}) \land$ \\
 &  & $|w_j|~(e_j \lor \neg b_{2j} \lor \neg b_{2j+1})$ \\
\hline
\multirow{4}{*}{\shortstack{2. If $w_j \geq 0$: negated literal \\ Note: To achieve a 3-SAT form we need 2 \\ auxiliary variables. Let's call them $b_{2j}$ and $b_{2j+1}$}} & \multirow{4}{*}{$w_j~\neg e_j$} & $w_j~(\neg e_j \lor b_{2j} \lor b_{2j+1}) \land$ \\
 &  & $w_j~(\neg e_j \lor \neg b_{2j} \lor b_{2j+1}) \land$ \\
 &  & $w_j~(\neg e_j \lor b_{2j} \lor \neg b_{2j+1}) \land$ \\
 &  & $w_j~(\neg e_j \lor \neg b_{2j} \lor \neg b_{2j+1})$ \\
\hline
\hline
\end{tabular}
\end{adjustbox}
\caption{Prescriptions for translating different types of equations or soft clauses to 3-SAT (weighted) CNF form.}
\label{tab:prescription}
\end{table*}

For the proposed mapping, we have $\omega_i - 2$ number of constraints in~\eqref{eq:xor_breakdown} and each of them would translate to 4 clauses as described in Tab.~\ref{tab:prescription}. Thus, the total number of clauses is $4\sum_i(\omega_i-2)$.

\subsection{Soft Clauses}

Next, we deal with finding the most likely error under the syndrome constraints, which is described by the maximisation condition~\eqref{eq:soft}.

MaxSAT algorithms allow for introducing a weight associated with each soft clause, which would be the cost of violating that particular clause. In this context, hard clauses are considered as infinite-weight clauses, which means that even if one of the hard clauses is violated, the solver has failed to find a satisfying assignment.

In the language of MaxSAT, the log-likelihood maximisation in~\eqref{eq:soft} can be described as a set of weighted soft clauses. One can drop the negative sign in front of the log-likelihood in~\eqref{eq:soft} and then the weights $w_i$ can be interpreted as costs of soft clauses to be minimised. Here, it is worth noting that the most likely error configuration is not necessarily the one with least Hamming weight, which was the assumption in~\cite{berent2023decoding}, although this is a valid assumption for uniform error probabilities.
Therefore, for probabilities $p_j<0.5$, the corresponding MaxSAT soft clause is the negation of the error literal $e_i$ with corresponding weight $w_j$ as its cost, as shown in Tab.~\ref{tab:prescription}, while for $p_j \geq 0.5$ a slight modification is necessary. 

To turn the soft clauses into 3-SAT clauses, we require $2n$ auxiliary literals $\{b_1, \cdots, b_{2n}\}$ and the total number of soft clauses would be $4n$; see Tab.~\ref{tab:prescription}.

We stress that in our MaxSAT formulation, the error probabilities $p_j$ for each data qubit $j$ can be chosen arbitrarily. Further, our mapping applies to parity check matrix of arbitrary CSS codes. 
We note that the MaxSAT decoder of Ref.~\cite{berent2023decoding} only supports uniform $p_j$, and is restricted to the color code.

\subsection{Measurement error}
So far, we considered the decoding problem where we assume that syndrome measurements are not affected by noise. 
We now handle noisy syndrome measurements via a phenomenological error model~\cite{dennis2002topological}. 
Here, we assume that the syndrome outcome of the $i$th parity check measurement is flipped with probability $q_i$. Now, a non-zero syndrome outcome may indicate that either an error occurred on the data qubits or on the measurement itself. 
To be able to pinpoint the origin of the error, one repeats the syndrome measurement $L$ times (with usually $L=d$), and pinpoints the error by performing parity checks  across both the data qubits and consecutive noisy syndromes.
As before, we can decode $X$ and $Z$ errors separately, where we now illustrate decoding $H_Z$ for $X$ errors.

We denote the index of measurement repetition by $t=1,\dots,L$. Errors on the data qubits that happen during the $t^\text{th}$ iteration are given by $\boldsymbol{e}^t=\{e^t_1,\dots,e^t_n\}$ where each error happens with probability $p_j$. Additionally, errors in the syndrome measurement during the $t^\text{th}$ measurement is described by the binary list $\boldsymbol{r}^t=\{r^t_1,\dots,r^t_m\}$. 
The quantum maximum likelihood decoding problem can be written as
\begin{align}
   (\boldsymbol{e}^{\text{ml}}, \boldsymbol{r}^{\text{ml}}) &= \arg \max_{\boldsymbol{e}, \boldsymbol{r}}  \{\ln(P(\boldsymbol{e},\boldsymbol{r}))\} \nonumber \\
   &=\arg \max_{\boldsymbol{e}, \boldsymbol{r}} (C'-\sum_{j=1}^n\sum_{t=1}^L w_j e_j^t-\sum_{i=1}^m\sum_{t=1}^{L-1} u_i r_i^t ) \label{eq:soft_meas_error}\\
    \text{s.t. }\,  \boldsymbol{s}'&=H_Z'\boldsymbol{v} \,,\label{eq:hard_meas_error}
\end{align}
where $u_i=\ln((1-q_i)/q_i)$ are the weights for the syndrome error probability, $\boldsymbol{s}'=\{\boldsymbol{s}^1,\boldsymbol{s}^1\oplus\boldsymbol{s}^2,\dots,\boldsymbol{s}^{L-1}\oplus\boldsymbol{s}^L\}$ the difference between consecutive syndrome outcomes with noisy syndrome $\boldsymbol{s}^t\in\{0,1\}^m$ measured at the $t^\text{th}$ repetition, error configuration vector $\boldsymbol{v}=\{\boldsymbol{e}^1,\boldsymbol{r}^1,\boldsymbol{e}^2,\boldsymbol{r}^2,\dots,\boldsymbol{e}^{L-1},\boldsymbol{r}^{L-1},\boldsymbol{e}^L\}$ combining data and measurement error literals, and the parity check matrix including measurement errors $H_Z'$ (which is a $mL\times (nL+m(L-1))$ matrix) given by
\begin{equation*}
    H_Z'=\begin{pmatrix}
        H_Z &I_{m} &0 &0 &0 &0&\dots &0&0&0& 0\\
        0 &I_{m} &H_Z &I_{m} &0 &0&\dots &0&0&0& 0\\
        0 &0 &0 &I_{m} &H_Z &I_{m}&\dots &0&0& 0& 0\\
        0 &0 &0 &0 &0 &I_{m}&\dots &0 &0&0& 0 \\
        \vdots &\vdots &\vdots &\vdots &\vdots &\vdots&\ddots &\vdots &\vdots&\vdots&\vdots\\
        0 &0 &0 &0 &0 &0 &\dots & I_{m}& 0 & 0 &0\\
        0 &0 &0 &0 &0 &0 &\dots & I_{m}& H_Z & I_{m} &0\\
        0 &0 &0 &0 &0 &0 &\dots & 0& 0 & I_{m} &H_Z
    \end{pmatrix}
\end{equation*}
with $I_{m}$ being the $m\times m$ identity matrix.
Note that one can assume that the measurement of the last syndrome is noise-free.
Now, each parity checks of $H_Z'$ (i.e. each row)  has $\omega_i+2$ non-zero terms (except in the first and last repetition), where $\omega_i$ terms interact with data qubit errors and $2$ terms with measurement errors.
Checks related to the first and last repetition have only $\omega_i+1$ terms: This is because the first check has no preceding error syndrome, while for the last repetition we assume that the last syndrome measurement is noise-free. 
Note that for the case $L=1$, we recover the noiseless syndrome decoding problem in~\eqref{eq:soft} and \eqref{eq:hard}.

Similar to the noise-free syndrome measurement case, the decoding problem including measurement noise can be mapped into a MaxSAT instance. The key point to note is that all $L$ rounds of the syndrome measurements for every $m$ parity checks of $H_Z$ are decoded simultaneously as implied by the extended parity check matrix $H_Z'$ in \eqref{eq:hard_meas_error}. The corresponding hard constraint equations for all $mL$ parity checks (indexed below by $i$) is given consistently by (similar to \eqref{eq:constraint_i})
\begin{equation}\label{eq:constraint_i_meas_error}
   s'_i = h'_{i,1}v_1 \oplus h'_{i,2}v_2 \oplus \cdots \oplus h'_{i,(nL+m(L-1))}v_{(nL+m(L-1))} \,.
\end{equation}
where we used the previously introduced vector of error variables $\boldsymbol{v}$, syndrome differences $\boldsymbol{s}'$ and matrix elements $h'_{ij}$ of $H'_Z$. The breakdown of equations in \eqref{eq:constraint_i_meas_error} into smaller ones for mapping to 3-SAT clauses is analogue to \eqref{eq:xor_breakdown}, which we avoid repeating here for brevity. The Tab.~\ref{tab:prescription} contains all the necessary and sufficient rules to obtain the 3-SAT form of the \textit{XOR} equations for this case.

The soft clauses is constructed similarly as before with now two weight vectors, namely $w_j$ for the qubit literals and $u_i$ for syndrome literals.

\section{Clause density and complexity}\label{sec:clausedens}
A key factor in complexity of solving MaxSAT instances is the clause density $\alpha$. In the study of computational complexity, the clause density of a random \textit{k}-SAT problem is defined as the ratio of the number of clauses \( \mu \) to the number of literals \( \nu \), denoted by \( \alpha = \mu/\nu \). The hardness phase transition for \textit{k}-SAT problems is a critical concept, highlighting a threshold clause density $\alpha_c$ at which the problems transition from being predominantly solvable to predominantly unsolvable. If we represent the class of random $k$-SAT problems with $\nu$ literals and $\mu$ clauses as $\mathcal{R}_k(\nu, \mu=\alpha \nu)$, then the hardness phase transition can be written as the discontinuity in the success probability of finding a satisfying assignment to the problems as a function of clause density (the order parameter) \cite{sat_phase_Gent}
\begin{equation}
    \lim_{\nu \to \infty} \text{Prob}(sat, \mathcal{R}_k(\nu, \alpha \nu)) =
\begin{cases} 
    \text{0}, & \text{if} ~ \alpha > \alpha_c\\
    \text{1}, & \text{if} ~ \alpha < \alpha_c
\end{cases}
\end{equation}
For different values of \( k \), the critical clause density where this phase transition occurs varies. For \( k = 2 \), the phase transition occurs at $\alpha_\text{c} \approx 1.0$~\cite{sat_phase_Gent}. For \( k = 3 \), the phase transition occurs at $\alpha_\text{c} \approx 4.2$~\cite{sat_phase_Gent}, a widely studied and critical threshold in the theory of NP-completeness. For \( k = 4 \), the phase transition is observed at \( \alpha \approx 9.8 \)~\cite{sat_phase_Gent}. 
The direct relevance of these transitions in the context of QEC decoding, is in the scaling of the time to solution. For example, it has been shown that for random 3-SAT instances at $\alpha_c > 5.6$, resolution algorithms need \textit{exponential time} to get a probability of success near 1~\cite{resol_alg_chvatal, sat_phase_Gent}. 
Therefore, the study of the complexity phase transition provides a theoretical framework for expecting a sudden shift in the solvability of the problems as the clause density crosses these thresholds. For a detailed discussion see also Ref.~\cite{bresler2021algorithmic}. Therefore, the critical clause density can have important implications on whether it is feasible to use SAT solvers as QEC decoders and which QEC codes would be ideal for these solvers.

Apart from the hardness phase transition as a factor to consider in choosing the clause length, it is also important to note that lower clause lengths are more desirable. This is due to the fact that designing a hardware SAT solver for lower $k$ values is easier. In other words, one can think of a clause as a $k$-body interaction term where we opt to use $3$-body terms.  

Optimally we would like the CNF of the decoding problem to be in the easy phase $\alpha < \alpha_c $. 

Now we bound the clause density for generic CSS codes (assuming no syndrome measurement errors). Assuming the $n$-qubit code has $m$ parity checks each with Hamming weight $\omega_i$, then for the 3-SAT construction in Tab.~\ref{tab:prescription} the clause density is given by
\begin{equation}\label{eq:alpha_3sat}
    \alpha^{\texttt{3-SAT}} = \frac{\#~clauses}{\#~literals} = \frac{\sum_{i=1}^m 4(\omega_i - 2)}{n + \sum_{i=1}^m (\omega_i - 1)}
\end{equation}
For common CSS codes such as the Toric code, one has $m\approx n/2$. To get a simple formula for $\alpha$, let us consider $m\leq n$. Therefore, we get $\alpha \leq 4 - \frac{8}{\omega}\leq 4$. In particular, the clause density always remains below the phase transition $\alpha_\text{c}\approx 4.2$, which holds in fact for any $m$.

Similarly, one can also consider the clause density for the MaxSAT decoding including the soft clauses that encode the error probability.
For a homogeneous Max-3-SAT CNF formula, the clause density is
\begin{equation}\label{eq:alpha_max_3sat}
    \alpha^{\texttt{Max-3-SAT}} = \frac{\#~clauses}{\#~literals} = \frac{4n + \sum_{i=1}^m 4(\omega_i - 2)}{3n + \sum_{i=1}^m (\omega_i - 1)}
\end{equation}
Addition of soft clauses in fact reduces the clause density, giving us again an upper bound of $\alpha\leq 4$ for any $m$.
It was shown that critical value of the clause density at the phase transition point for random Max-k-SAT instances is similar to the random k-SAT ones~\cite{max_k_sat_transition}.

As an example, we now estimate $\alpha^{\texttt{3-SAT}}$ for the triangular $6.6.6$ color code used in \eqref{eq:hard}.
This code is on a lattice with triangular geometry and hexagonal lattice tiles, where tiles are also referred to as faces or parity checks.
In such a lattice, the distance of the code $d$ is the linear dimension of the lattice (length of one side), and the total number of qubits present is proportional to $\frac{3d^2 + 1}{4}$. The total number of faces (parity checks) is also proportional to $\frac{3d^2 + 1}{8}$. As explained in previous sections, we need $4(\omega_i-2)$ 3-SAT clauses for $i^\text{th}$ parity check. For the considered color code, $\omega_i \in \{4, 6\}$. Since the majority of the parity checks have weight $6$, we pick $\omega_i = \omega = 6$ to estimate an upper-bound for the clause density. Therefore the total number of clauses is proportional to $4(\omega-2)\frac{3d^2 + 1}{8} \propto 6d^2$. Now we need to estimate the number of literals used in each instance. It is straightforward to show that the number of literals is proportional to $\frac{3d^2 + 1}{4} + \frac{3d^2 + 1}{8} (\omega-1) \propto 2.6 d^2$. Therefore, the clause density is $\alpha^{\texttt{3-SAT}} = \frac{6d^2}{2.6d^2} \approx 2.3$ which is significantly less than $\alpha_c = 4.2$. Thus, with the proposed construction the color code is deeply within the \textit{easy} phase of MaxSAT even when scaling up the code size.

The clause density calculation for the case including noisy syndrome measurements follows similar to~\eqref{eq:alpha_3sat} and~\eqref{eq:alpha_max_3sat}. The main differences are the number of parity checks which have increased by $L$ times for $H'_Z$ and each of them now involve $4\omega_i$ 3-SAT clauses (except for $i \in \{1, mL\}$ where there are $4(\omega_i-1)$ clauses) and additionally the number of error variables has increased to $nL$ qubit error variables represented by $\boldsymbol{e}$ and $m(L-1)$ syndrome error variables represented by $\boldsymbol{r}$ (see \eqref{eq:soft_meas_error}). 
Finally, for soft clauses, we need to add the contribution of the qubit and syndrome error variables. For qubit error variables we get a total of $4nL$ soft clauses and $2nL$ auxiliary variables whereas for syndrome error variables we get $4m(L-1)$ additional soft clauses and $2m(L-1)$ additional auxiliary literals. Therefore the clause density for the hard constraints is
\begin{equation}\label{eq:synd_alpha_3sat}
    \alpha^{\texttt{3-SAT}} = \frac{\sum_{i=1}^{mL} 4\omega_i}{nL + m(L-1) + \sum_{i=1}^{mL} (\omega_i + 1)}
\end{equation}
and for MaxSAT with both soft and hard constraints is given by
\begin{equation}\label{eq:synd_alpha_max_3sat}
    \alpha^{\texttt{Max-3-SAT}} = \frac{4nL + 4m(L-1) + \sum_{i=1}^{mL} 4\omega_i}{3nL + 3m(L-1) + \sum_{i=1}^{mL} (\omega_i + 1)}
\end{equation}
As an example, for the 6.6.6 color code, we get $\alpha^{\texttt{Max-3-SAT}}\approx 2.5$. We note that again $\alpha^{\texttt{Max-3-SAT}}\leq4$, i.e. the MaxSAT problem always remains in the \emph{easy} computational phase. 

\section{Methods}\label{sec:methods}
To solve MaxSAT, we assessed the performance of various MaxSAT solvers and  the best performing algorithm was chosen to carry out the experiments. Based on qualitative and quantitative results, among various solvers such as CASHWMaxSAT-CorePlus, CGSS2, EvalMaxSAT, Open-WBO, Pacose and Loandra~\cite{maxsat_eval}, Open-WBO proved to be the best solver.

To evaluate the performance of our MaxSAT decoder, we compare against another popular decoder that runs on arbitrary CSS codes, namely belief propagation with the
ordered statistics postprocessing step decoder (BP-OSD) which was proposed in Ref.~\cite{roffe2020decoding,panteleev2021degenerate}. 
BP-OSD relies on belief propagation, a heuristic message passing algorithm that estimates the single-bit marginals of a probability distribution, with a ordered statistics postprocessing step to determine the most likely error.
BP-OSD has been shown to work well for bivariate bicycle QLDPC codes~\cite{kovalev2013quantum} such as the codes proposed in Ref.~\cite{Bravyi_2024}. 

For both MaxSAT and BP-OSD, we decode $H_X$ and $H_Z$ separately, giving us the total decoded qubit error $\boldsymbol{e}_\text{dec}=\bigoplus_{t=1}^L\boldsymbol{e}^t$. We compute the remaining error after correction $\delta\boldsymbol{e}=\boldsymbol{e}^\ast\oplus\boldsymbol{e}_\text{dec}$ by adding the decoded data qubit error onto the actual error configuration $\boldsymbol{e}^\ast$.  The logical error rate $p_\text{L}$ is given by the probability that after correction the remaining error $\delta\boldsymbol{e}$ anti-commutes with at least one of the logical operators of the $k$ logical qubits.

As error model, we assume single-qubit depolarizing noise
\begin{equation}
    \mathcal{E}_p(\rho)=(1-p)\rho +p/3( X\rho X+ Y\rho Y+ Z\rho Z)
\end{equation}
with physical error rate $p_j^X=p_j^Z=p_j^Y=p/3$ which are uniform for every qubit $j$. Further, we assume noise-free syndrome measurements $q_j=0$ and one repetition $L=1$, i.e. we study the code capacity.

To characterize the error capabilities of our codes, we study the scaling of the logical error logical error $p_\text{L}$ against physical error rate $p$ with the heuristic formula~\cite{bravyi2013simulation, Bravyi_2024}
\begin{equation}
    p_\text{L}(p)=p^{ d_\text{fit}/2 }\exp(c_0+c_1p+c_2p^2)\,.
\label{eq:heuristic_fit}
\end{equation}
For surface codes and bicycle codes~\cite{Bravyi_2024,voss2024trivariate} it has been observed that the fitted distance corresponds to the actual distance via $d_\text{fit}/2=\lceil d/2 \rceil$~\cite{bravyi2013simulation}.

We also study the pseudo threshold $p_{\text{p-th}}$. It is defined as the error probability satisfying the break-even relation~\cite{Bravyi_2024} 
\begin{equation}
    p_{\text{L}}(p_{\text{p-th}}) = 1-(1-p_{\text{p-th}})^k
\end{equation}
where the probability of a logical error is equivalent to the probability of at least one unencoded physical qubit suffering an error. 

\section{Results}\label{sec:result}

\begin{figure}[htbp]
	\centering	\subfigimg[width=0.235\textwidth]{a}{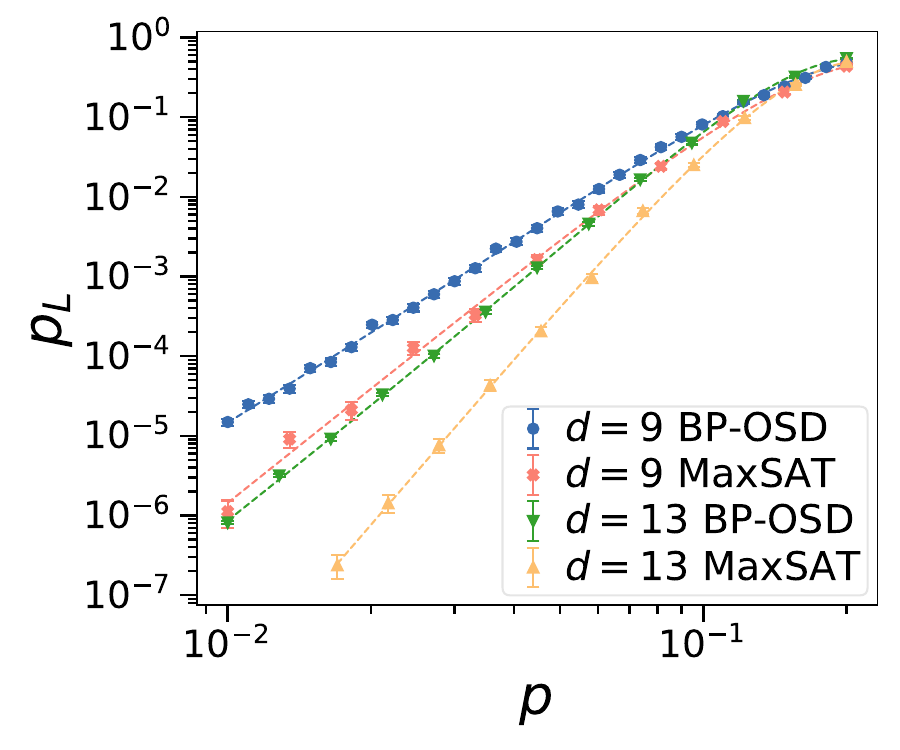}
 \subfigimg[width=0.235\textwidth]{b}{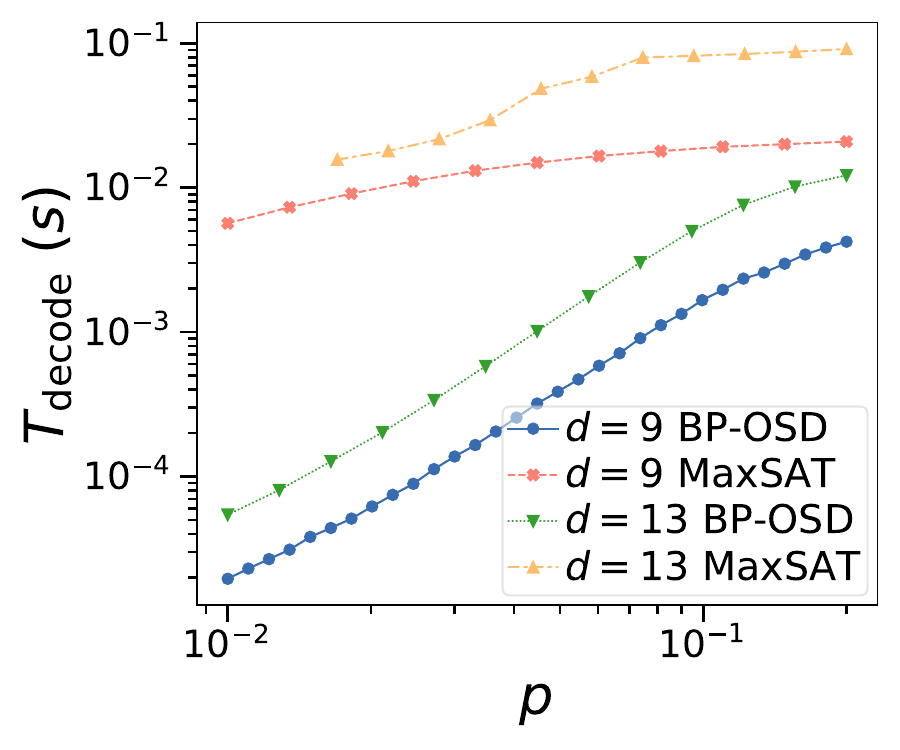}
	\caption{Performance of our MaxSAT decoder and BP-OSD decoder against physical error rate $p$ for the Color code. \idg{a} Logical error rate $p_\text{L}$  against $p$. Dashed line is fit with~\eqref{eq:heuristic_fit} with $d_\text{fit}^{\text{BP-OSD}}=7.3$ and $d_\text{fit}^{\text{MaxSAT}}=9.6$ for $d=9$, and $d_\text{fit}^{\text{BP-OSD}}=9.2$ and $d_\text{fit}^{\text{MaxSAT}}=13.8$ for $d=13$.
    We find pseudo-thresholds $p_\text{p-th}^{\text{BP-OSD}}=0.109$ and $p_\text{p-th}^{\text{MaxSAT}}=0.122$ for $d=9$, while $p_\text{p-th}^{\text{BP-OSD}}=0.112$ and $p_\text{p-th}^{\text{MaxSAT}}=0.130$ for $d=13$.
 \idg{b} Decoding time $T_\text{decode}$ against $p$. 
	}
	\label{fig:pcolor}
\end{figure}

First, we study the 6.6.6 or Honeycomb Color code~\cite{bombin2006topological} for our MaxSAT decoder and BP-OSD against distance $d$ of the code. Its code parameters are given by $\dsl (3d^2+1)/4,1,d\dsr$.
We show the logical error rate $p_\text{L}$ against physical error rate $p$ in Fig.~\ref{fig:pcolor}a. We study the MaxSAT decoder and BP-OSD for different distances $d$. We find that MaxSAT decoder has superior scaling in error suppression. In particular, from the fit with~\eqref{eq:heuristic_fit}, we find the optimal scaling $d_\text{fit}\approx d$ for the MaxSAT decoder. In contrast, for BP-OSD we only find a reduced $d_\text{fit}\approx 3/4 d$, indicating that the BP-OSD decoder can miss errors even with weight smaller than $d/2$.

\begin{figure}[htbp]
	\centering	\subfigimg[width=0.235\textwidth]{a}{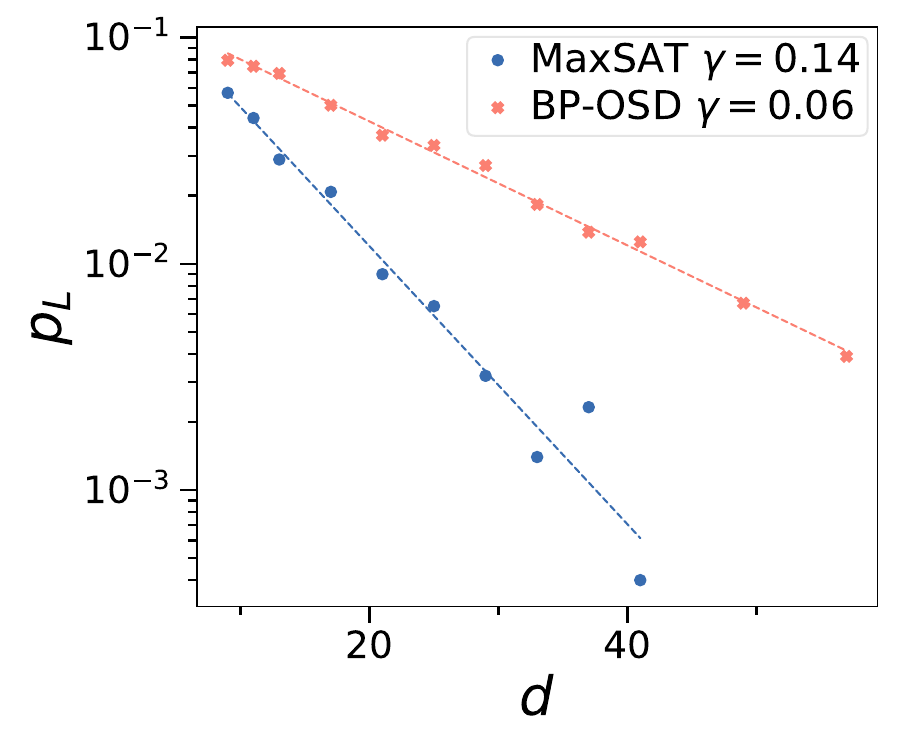}
 \subfigimg[width=0.235\textwidth]{b}{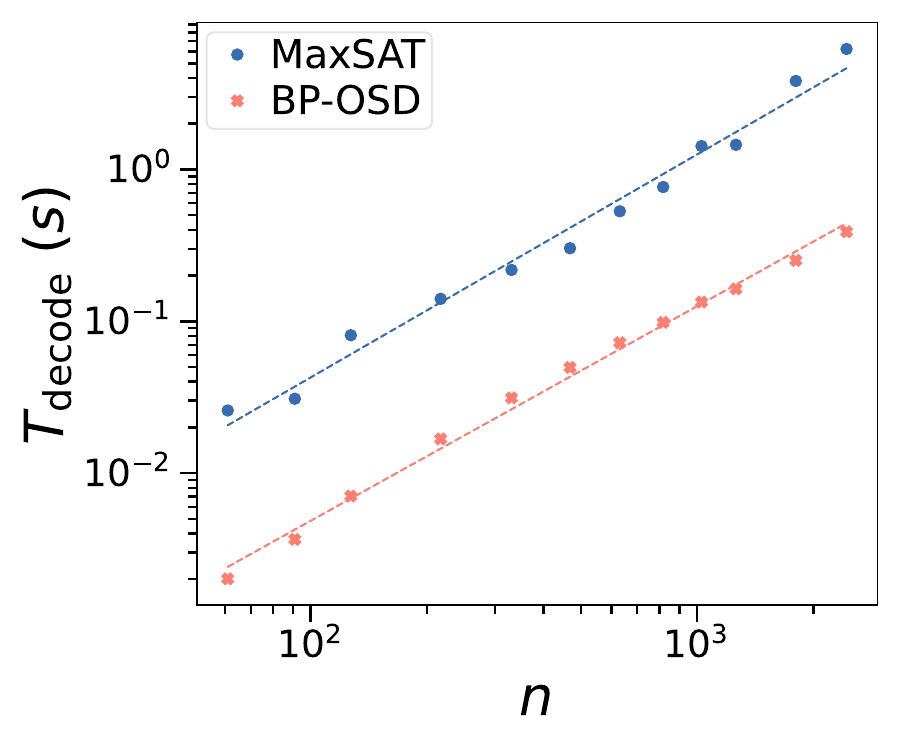}
	\caption{Scaling of MaxSAT and BP-OSD decoder for the Color code with error probability $p=0.1$. \idg{a} Logical error rate $p_\text{L}$  against distance $d$. Dashed line is fit with $p_\text{L}\propto \exp(-\gamma d)$ where we find $\gamma^{\text{MaxSAT}}\approx0.14$ and $\gamma^{\text{BP-OSD}}\approx0.06$.
 \idg{b} Decoding time $T_\text{decode}$ against number of data qubits $n$. Dashed line is fit with $T_\text{decode}\propto n^{\beta}$ with $\beta^{\text{MaxSAT}}\approx 1.46$ and $\beta^{\text{BP-OSD}}\approx 1.41$.
	}
	\label{fig:distancecolor}
\end{figure}

Next, we study $p_\text{L}$ against distance $d$ for fixed error $p=0.1$ in Fig.~\ref{fig:distancecolor}a. We find that our MaxSAT decoder has vastly lower logical error rate compared to BP-OSD. 
Both codes decrease the error exponentially $p_\text{L}\propto \exp(-\gamma d)$ with distance $d$. However, MaxSAT decoding shows a twice as fast decay $\gamma_\text{MaxSAT}\approx 0.14$ with $d$ compared to BP-OSD with $\gamma\approx0.06$. 
Thus, our MaxSAT decoder achieves the same noise suppression as BP-OSD at only half the code distance. Thus, our decoder yields a reduction of a factor $4$  in terms of physical qubit number compared to BP-OSD. We note that the reduced performance of BP-OSD for color codes has also been noted in Ref.~\cite{higgott2023improved}.

We show the time $T_\text{decode}$ needed to decode against number of data qubits $n$ in Fig.~\ref{fig:distancecolor}b. We find that MaxSAT decoder and BP-OSD decoder show similar scaling $T_\text{decode}\propto n^{\beta}$, with $\beta\approx 3/2$. We note that our MaxSAT decoder runs about an order of magnitude slower than BP-OSD. We note that our current implementation runs only on CPU, and an optimised implementation on specialized hardware can speed up the MaxSAT decoder substantially~\cite{sohanghpurwala2017hardware}.

\begin{figure}[htbp]
	\centering	\subfigimg[width=0.235\textwidth]{a}{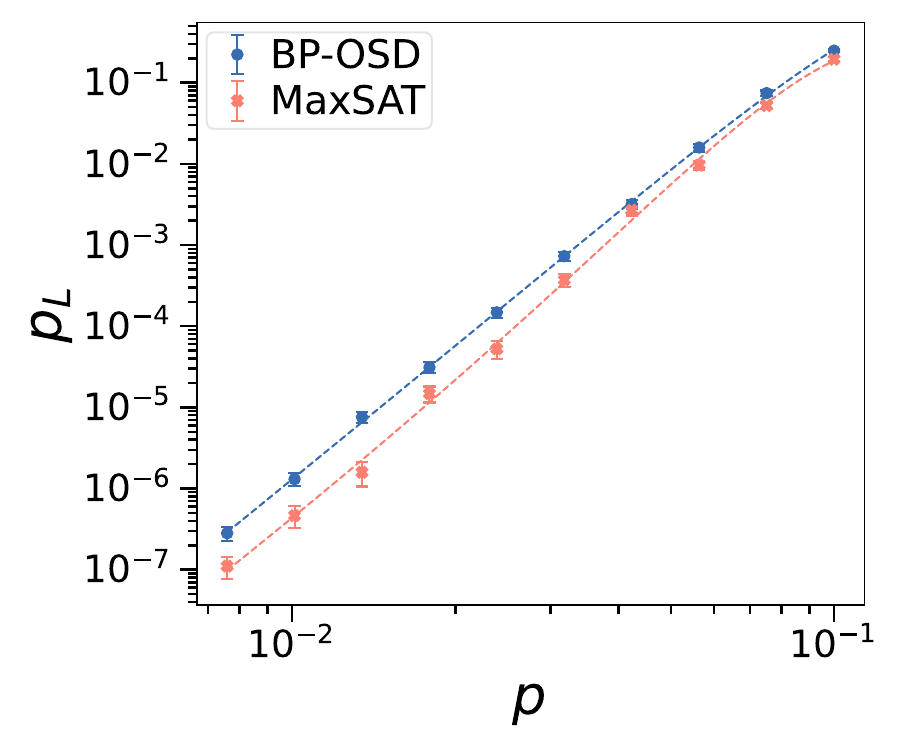}
 \subfigimg[width=0.235\textwidth]{b}{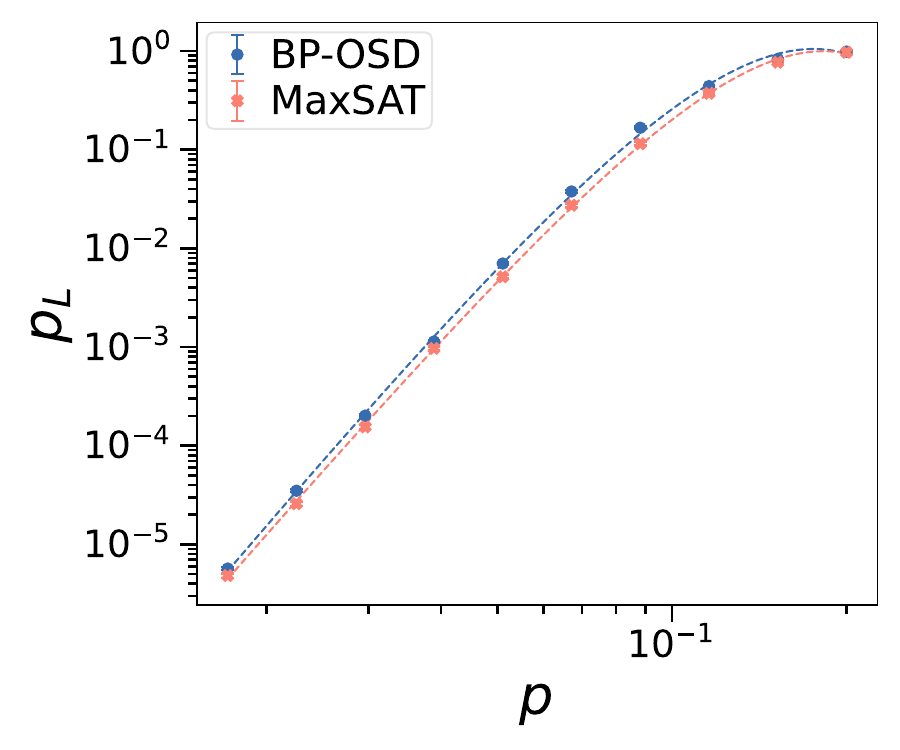}
	\caption{Performance for decoding the QLDPC codes introduced by IBM~\cite{Bravyi_2024} MaxSAT decoders and BP-OSD decoder.  We show logical error rate $p_\text{L}$ against physical error rate $p$. Dashed line is fit with~\eqref{eq:heuristic_fit}.
 \idg{a} $\dsl108,8,10\dsr$ QLDPC code with $d_\text{fit}^{\text{BP-OSD}}=10.7$, $d_\text{fit}^{\text{BP-OSD}}=9.8$.
 \idg{b} $\dsl144,12,12\dsr$ QLDPC code with $d_\text{fit}^{\text{BP-OSD}}=14.2$, $d_\text{fit}^{\text{BP-OSD}}=13.7$.
	}
	\label{fig:IBM}
\end{figure}

Next, in Fig.~\ref{fig:IBM} we study the decoding of  the QLPDC codes recently introduced by IBM~\cite{Bravyi_2024}. In Fig.~\ref{fig:IBM}a, we show a $\dsl108,8,10\dsr$ code, where we find that our MaxSAT decoder shows about three times lower logical error rates compared to the BP-OSD decoder~\cite{roffe2020decoding}, which was also used in Ref.~\cite{Bravyi_2024}. 
In Fig.~\ref{fig:IBM}b, we study the $\dsl144,12,12\dsr$ QLPDC code, where we find that our MaxSAT decoder has lower logical error rates compared to BP-OSD. 
We note that the logical error rate shows comparable scaling with $p$ for both MaxSAT and BP-OSD evident by similar values for $d_\text{fit}$.

\begin{figure}[htbp]
	\centering	\subfigimg[width=0.235\textwidth]{a}{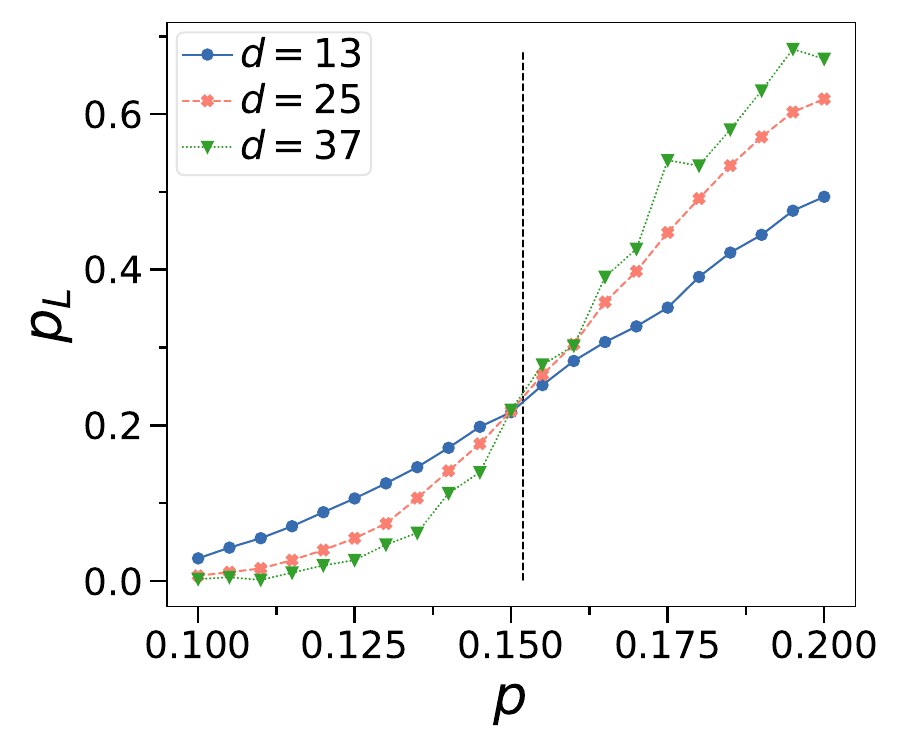}
\subfigimg[width=0.235\textwidth]{b}{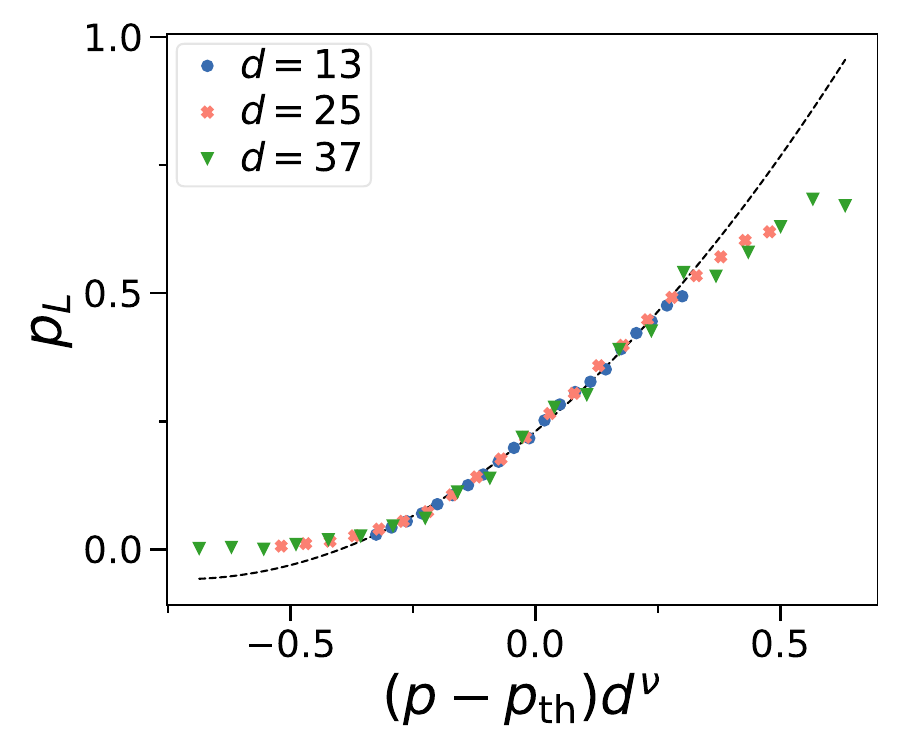}
 \subfigimg[width=0.235\textwidth]{d}{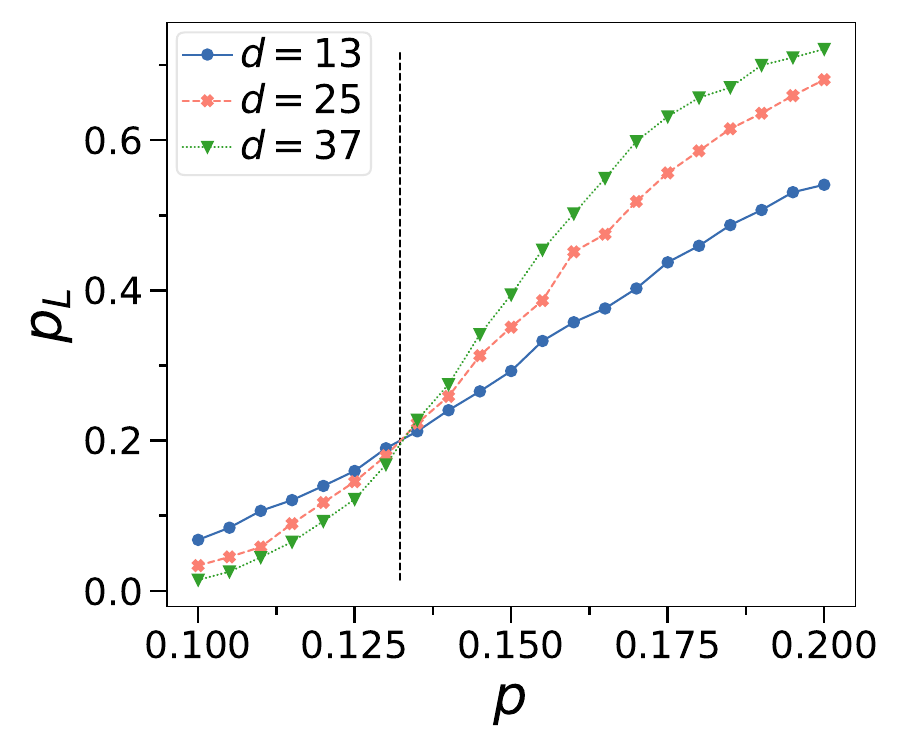}
 \subfigimg[width=0.235\textwidth]{d}{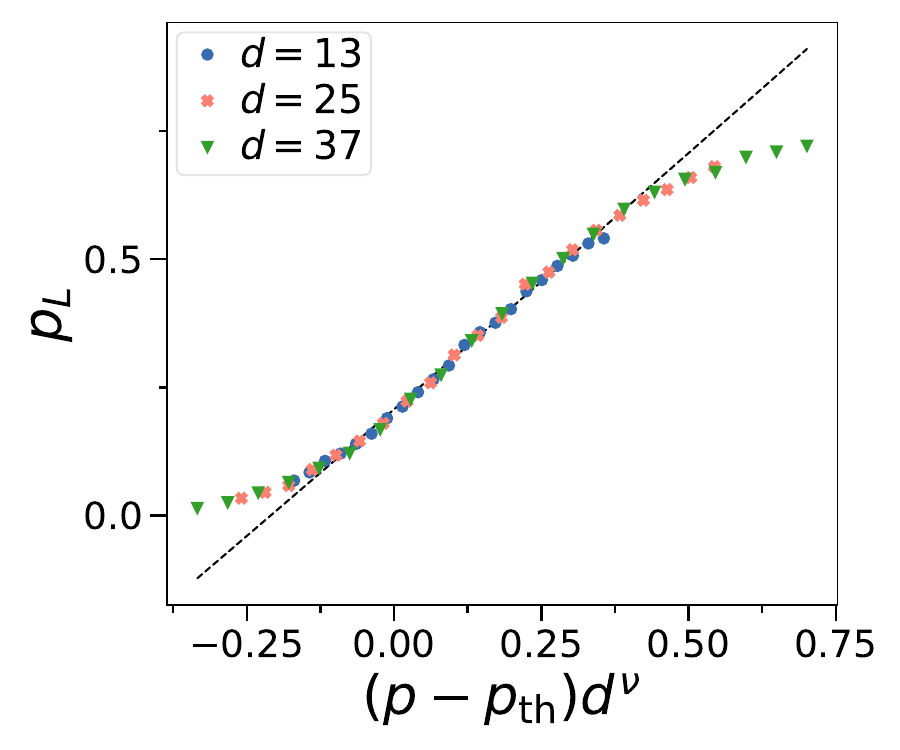}
	\caption{Threshold of Color code for  \idg{a,b} MaxSAT and \idg{c,d} BP-OSD decoder for depolarizing noise.
  \idg{a,c} We show logical error $p_\text{L}$ against physical error $p$ for different distances $d$ with fitted threshold $p_\text{th}=15.20\pm0.05\%$ for MaxSAT and $p_\text{th}=13.23\pm0.04\%$ for BP-OSD, indicated as dashed vertical line. 
  \idg{b,d}  Rescaled error around threshold $p_\text{th}$ of fit with  exponent $\nu=0.71$ for MaxSAT and $\nu=0.65$ for BP-OSD, where dashed line is the fitted critical scaling ansatz.
	}
	\label{fig:thresholdsColor}
\end{figure}

We study the threshold of the Color code in Fig.~\ref{fig:thresholdsColor} under depolarizing noise. We find that curves for different $d$ all intersect at the same physical error, which is the threshold $p_\text{th}$. We apply the critical scaling ansatz $p_\text{L}=f(d^\nu(p-p_\text{th}))$, with $f$ being a degree-$2$ polynomial, to fit $p_\text{th}$ and exponent $\nu$~\cite{chubb2021general}. When appropriately rescaled, curves for all $d$ collapse onto a single curve around the threshold. We find that the MaxSAT decoder has a higher fitted threshold of $p_\text{th}=15.20$ compared to the BP-OSD with $p_\text{th}=13.23$. 

\begin{figure}[htbp]
	\centering	
	\subfigimg[width=0.35\textwidth]{}{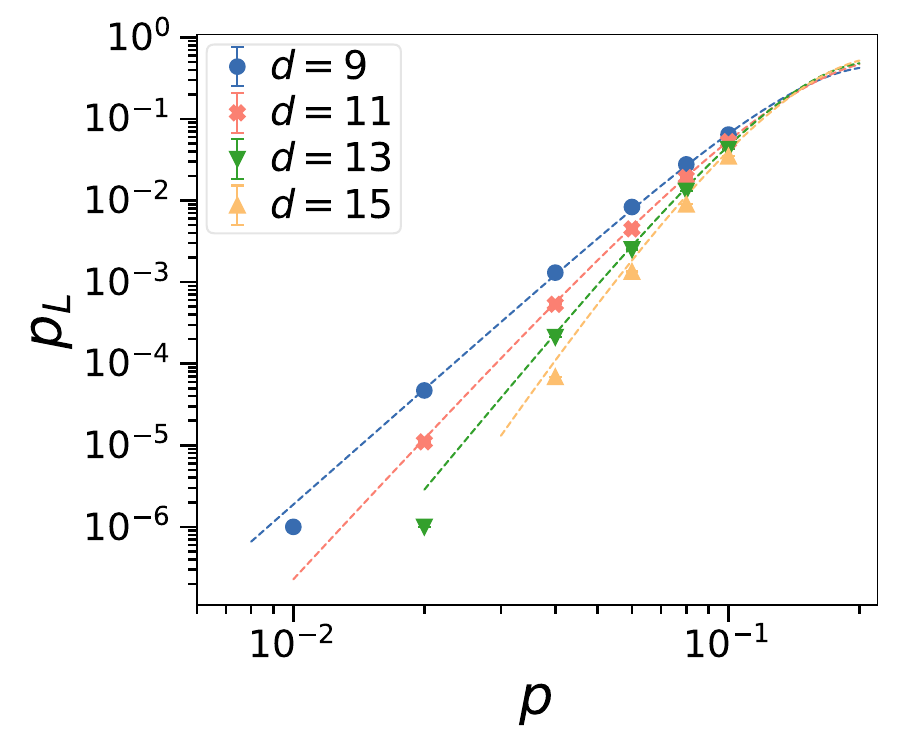}
	\caption{Logical error rate $p_\text{L}$ against physical error rate $p$ for the rotated surface code for different distances $d$. Dots are MaxSAT solver results, while dashed line is BP-OSD. 
	}
	\label{fig:rotatedscaling}
\end{figure}

Finally, we study the logical error rate $p_\text{L}$ against physical error $p$ for the rotated surface code~\cite{bombin2007optimal} in Fig.~\ref{fig:rotatedscaling}. We find that the MaxSAT decoder has slightly lower logical error rates compared to the BP-OSD decoder.

\section{Discussion}\label{sec:discussion}
We demonstrate MaxSAT to decode arbitrary CSS codes with high accuracy. 
We show how to map the quantum maximum likelihood decoding problem of CSS codes into Max-$3$-SAT instances. Our mapping supports CSS codes with parity checks of any weight and geometry, and can handle biased single-qubit Pauli noise with non-uniform error probabilities across different data qubits. Further, our decoder handles noisy syndrome measurements within the commonly used  phenomenological error model.

The computational complexity of solving MaxSAT is related to a computational phase transition at a critical clause density of the MaxSAT problem.
We show that the clause density of our mapping is always below the critical clause density, thus remaining in the \emph{easy} phase of MaxSAT. Thus, we expect our decoder to perform well even when scaling up the code size.

\begin{table}[!htbp]
    \centering
    \begin{tabular}{|c|c|c|c|c|c|}
        \hline
        \textbf{Code} & \textbf{MaxSAT} & \textbf{BP-OSD} \\
        \hline
        Toric Code (2D) & $15.55 \pm 0.03$ & $14.86 \pm 0.03$  \\
        \hline
        Color Code (on hexagonal lattice) & $15.20 \pm 0.05$ & $13.23 \pm 0.04$ \\
        \hline
    \end{tabular}
        \caption{Comparison of Error Thresholds (\%) for MaxSAT and BP-OSD decoders under uniform depolarizing noise.}
    \label{tab:error_thresholds}
\end{table}

We demonstrate that MaxSAT can decode the Color code effectively, showing a superior code capacity threshold compared to BP-OSD as summarized in Tab.~\ref{tab:error_thresholds}. Further, we find that MaxSAT decoding gives the asymptotically optimal scaling of $p_\text{L}\sim p^{d/2}$, while BP-OSD only reaches a sub-optimal scaling of $p_\text{L}\sim p^{\eta d/2}$ with $\eta\approx 3/4$.  From a scaling analysis in Fig.~\ref{fig:distancecolor}, our decoder allows a factor $4$ reduction in number of data qubits while offering similar noise protection than BP-OSD. 

We also study recently proposed QLDPC codes~\cite{Bravyi_2024}, finding improved accuracy compared to BP-OSD. 
Our solver also has a higher threshold for the Toric code compared to BP-OSD, and slightly lower error rates for the rotated surface code. 

We highlight that our MaxSAT decoder can be directly applied to arbitrary CSS codes as a quantum maximum likelihood decoder to find the most probable error. 
This contrasts more specialized decoders such as minimum-weight perfect matching (MWPM), which require that each error affects at most two decoders, i.e. each column of $H$ has at most two non-zero entries~\cite{dennis2002topological,fowler2013minimum,higgott2022pymatching}, and extensions to other codes require purpose-built modifications~\cite{lee2024color,sahay2022decoder,gidney2023new} or relaxations~\cite{wang2009graphical}. 

We find improved performance across different types QEC codes, and thus expect MaxSAT decoding to be robust to the specific implementation and application. In current work, while the decoder shows promising scaling with $n$, the solvers are still relatively slow in absolute computational time. However, we stress that we ran the MaxSAT solver serially on CPU, which leaves a lot of room for improvement in terms of speed-up. In particular, the solvers can be parallelised~\cite{schubert2010pamiraxt} and implemented on specialized hardware~\cite{sohanghpurwala2017hardware} such as 
Field Programmable Gate Arrays (FPGAs) and Application-specific integrated circuits (ASICs), which can result in speed-ups of several orders of magnitude. 
Thus, our approach promises application as practical and highly accurate decoders to enable quantum computing. 

\bibliography{references.bib}
\onecolumngrid

\appendix{}

\newpage
\section{Satisfiability Terminology}\label{appen:A}

In Tab.~\ref{tab:concepts}, we summarise concepts and conventions related to MaxSAT that were used in the paper.

\vspace{5mm}
\begin{tabular}{>{\bfseries}p{4cm}| p{12cm}}
    Boolean Variables & In \textit{k}-SAT, variables can take the values True or False. \\
    \hline
    Literals & A literal is a variable or its negation. For example, if \( x_i \) is a variable, then \( x_i \) and \( \neg x_i \) (not \( x_i \)) are literals. \\
    \hline
    Clauses & A clause is a disjunction (OR) of literals. In \textit{k}-SAT, each clause has exactly $k$ literals. For instance, \( (x_1 \lor \neg x_2 \lor x_3) \) is a 3-SAT clause. \\
    \hline
    Conjunctive Normal Form (CNF) & A Boolean formula is in CNF if it is a conjunction (AND) of clauses. For example, \( (x_1 \lor x_2 \lor \neg x_3) \land (\neg x_1 \lor x_2 \lor x_3) \) is a CNF formula. \\
    \hline
    Satisfiability & The \textit{k}-SAT problem asks whether there exists an assignment of True or False to the variables such that the entire CNF formula evaluates to True. 
    \label{tab:concepts}
\end{tabular}

\section{Extended Data}
\begin{figure*}[htbp]
	\centering	
	\subfigimg[width=0.35\textwidth]{a}{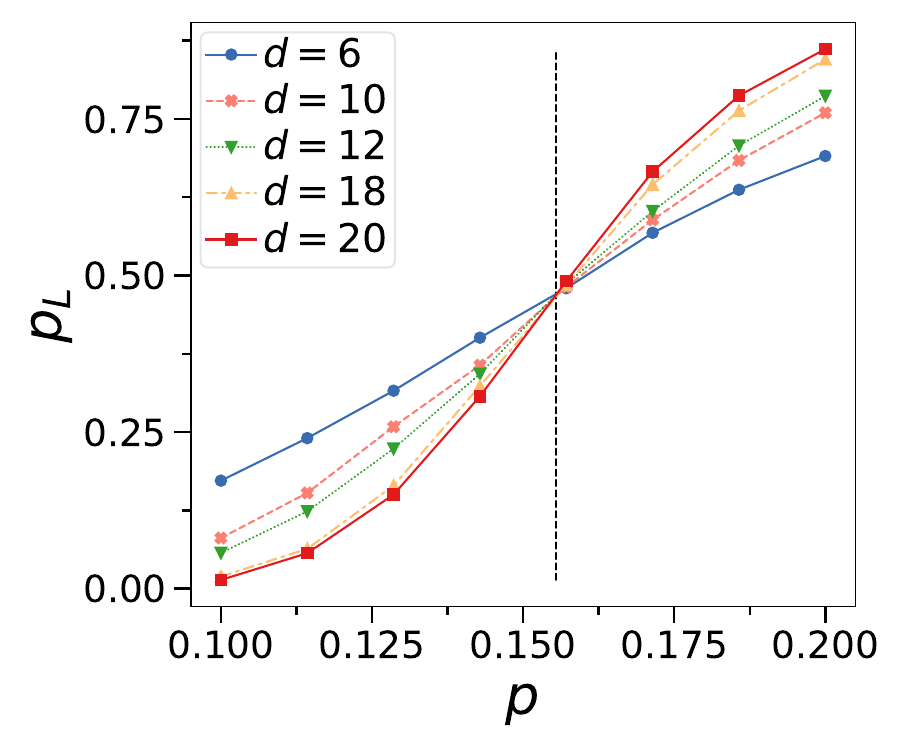}
 	\subfigimg[width=0.35\textwidth]{b}{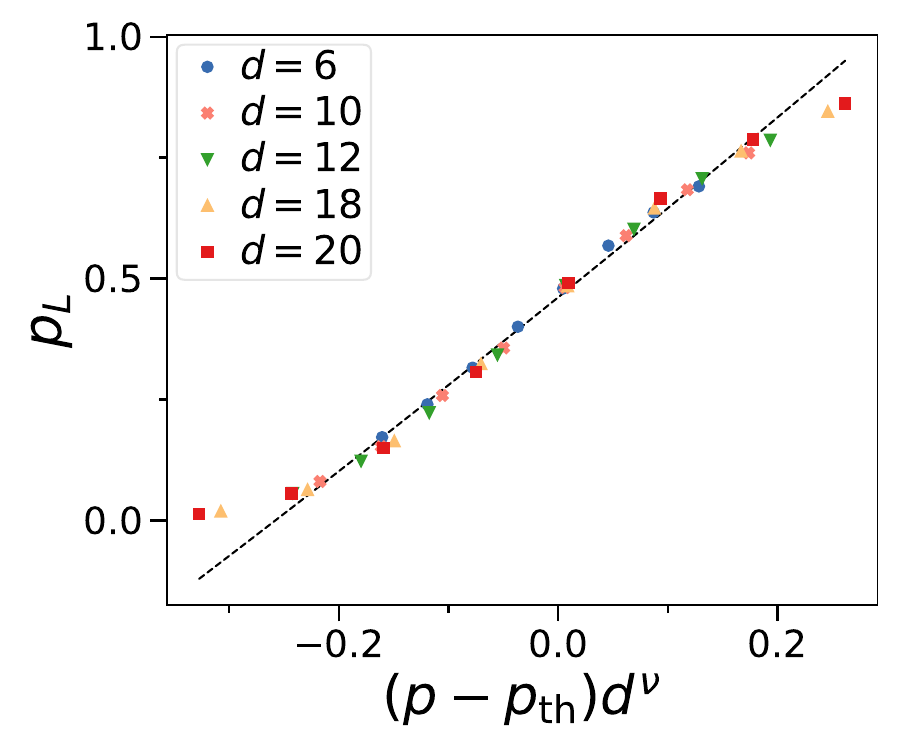}\\
	\subfigimg[width=0.35\textwidth]{c}{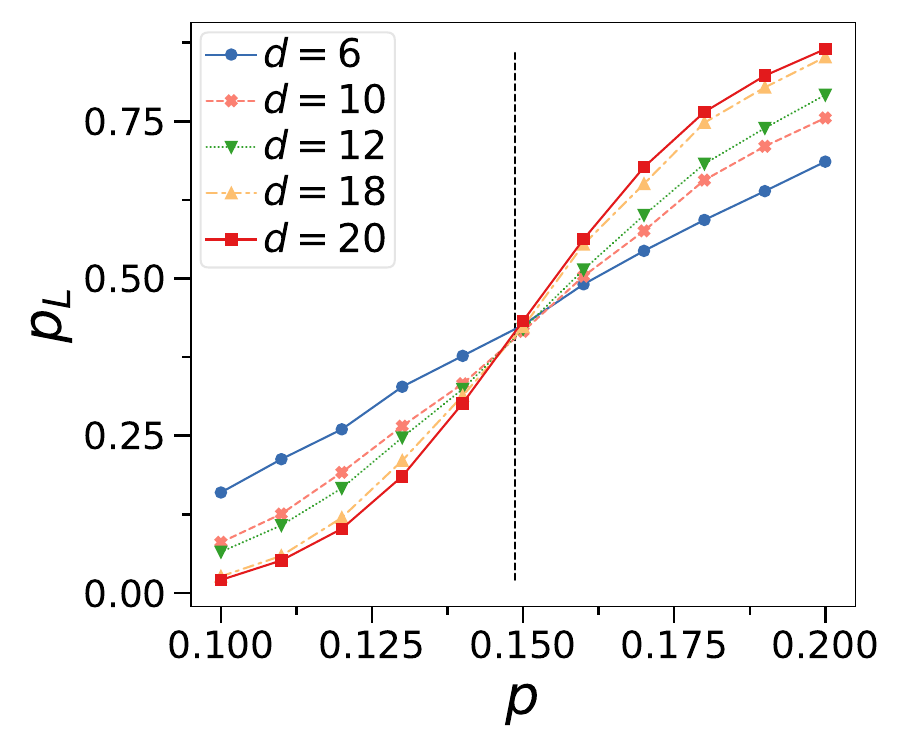}
 	\subfigimg[width=0.35\textwidth]{d}{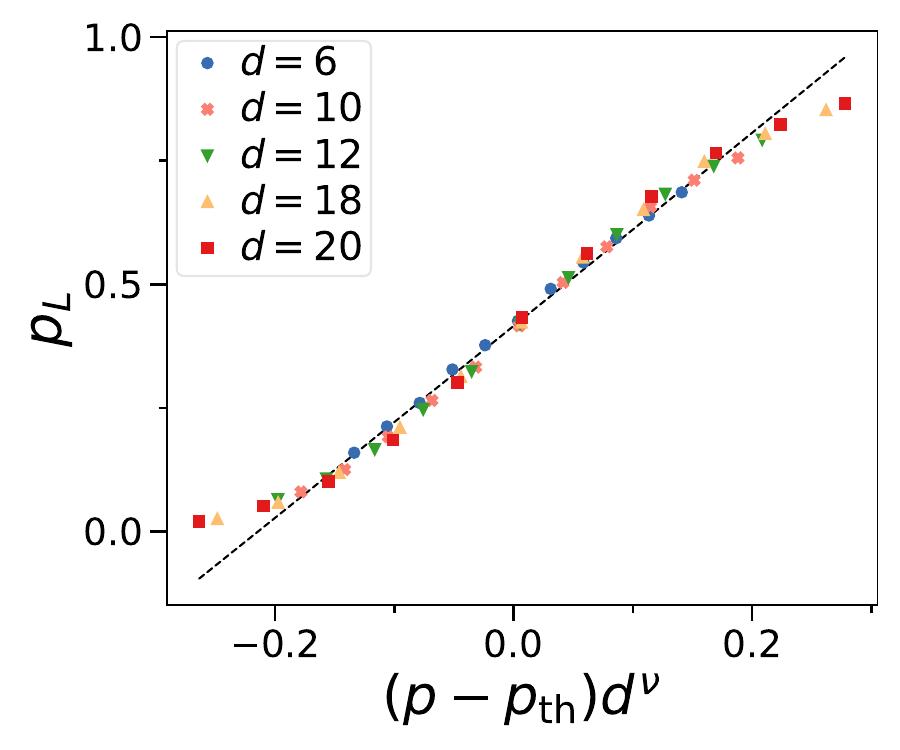}
	\caption{Threshold for the Toric code using our \idg{a,b} MaxSAT solver and \idg{c,d} BP-OSD. \idg{a,c} We plot logical error rate $p_\text{L}$ against physical error rate $p$ for depolarizing noise for different code distances $d$. We find threshold $p_\text{th}=15.55\pm0.03\%$ for MaxSAT and $p_\text{th}=14.86\pm0.03\%$ for BP-OSD. 
 \idg{b,d} Rescaled error plot around $p_\text{th}$  with exponent $\nu=0.59$ for MaxSAT and $\nu=0.56$ BP-OSD, where the dashed line is the fitted critical scaling ansatz.
	}
	\label{fig:Toric}
\end{figure*}

\begin{figure*}[htbp]
	\centering	
	\subfigimg[width=0.35\textwidth]{a}{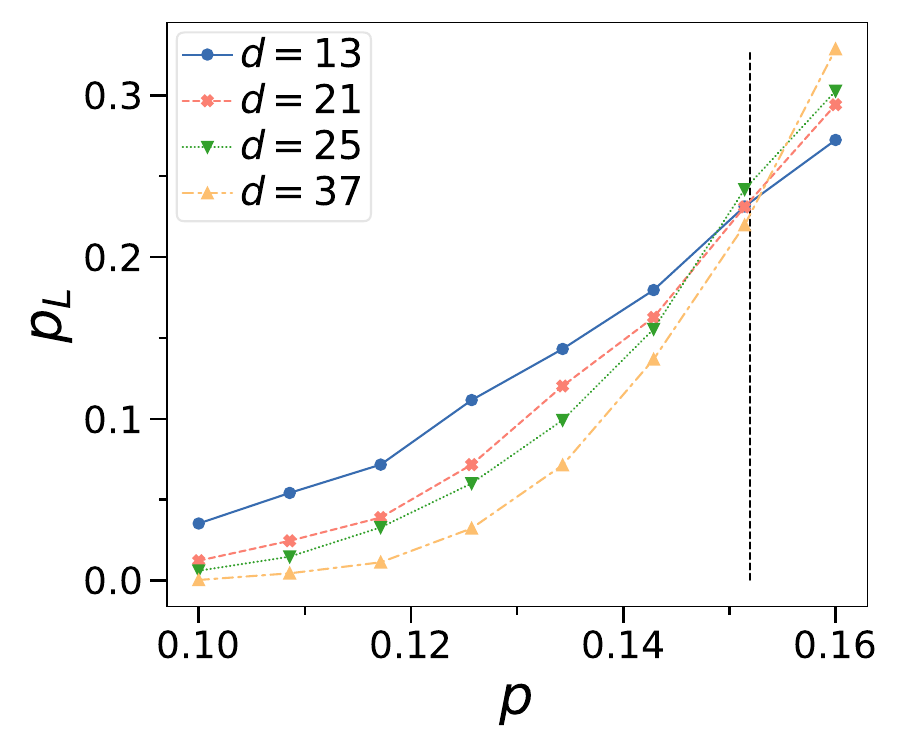}
 	\subfigimg[width=0.35\textwidth]{b}{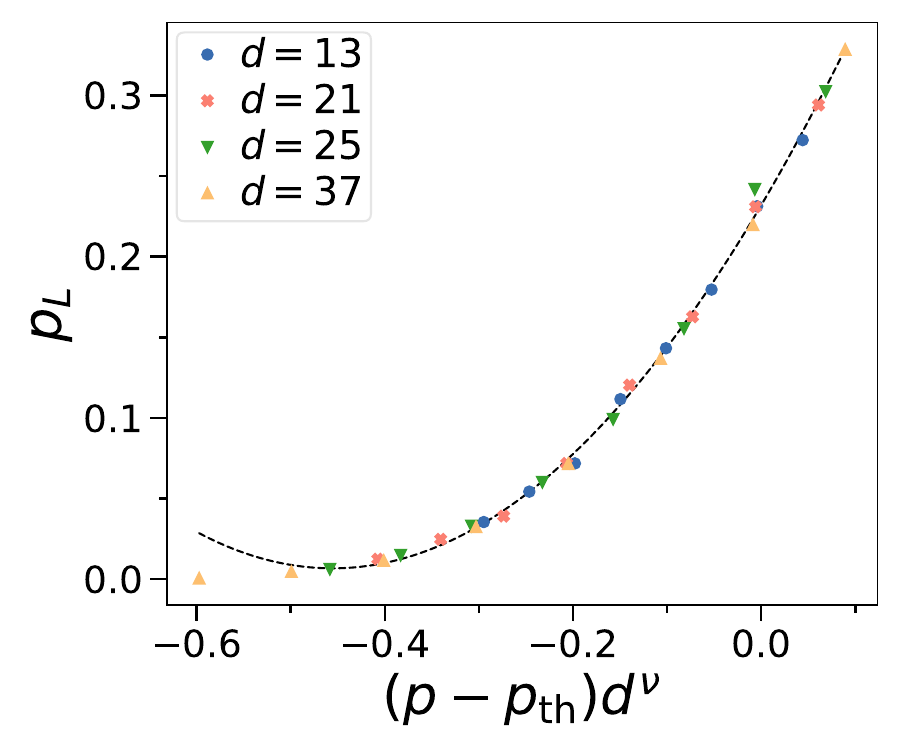}\\
	\subfigimg[width=0.35\textwidth]{c}{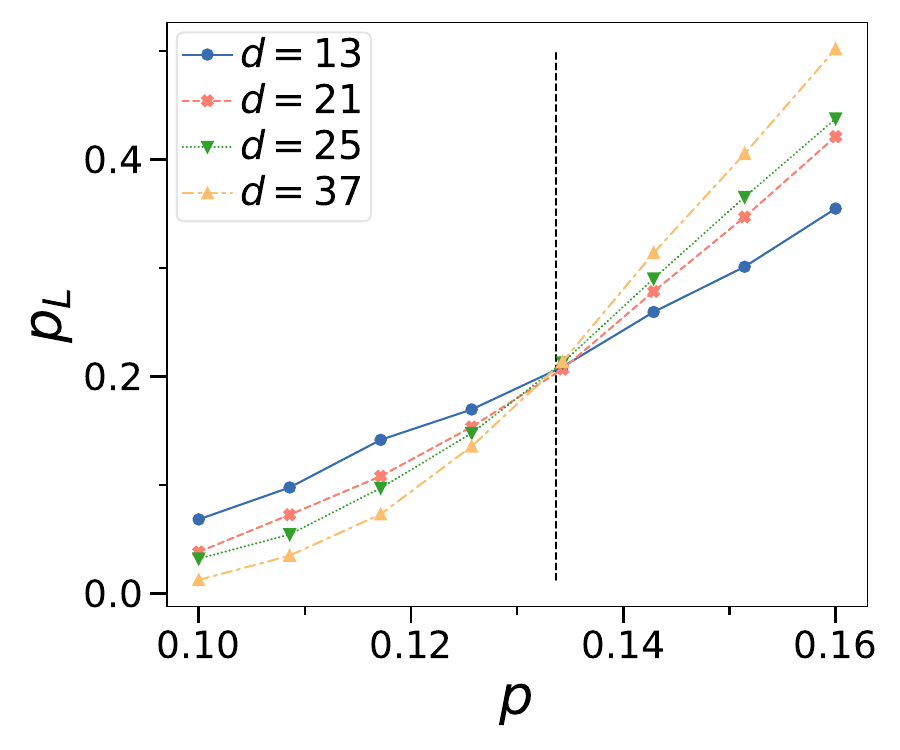}
 	\subfigimg[width=0.35\textwidth]{d}{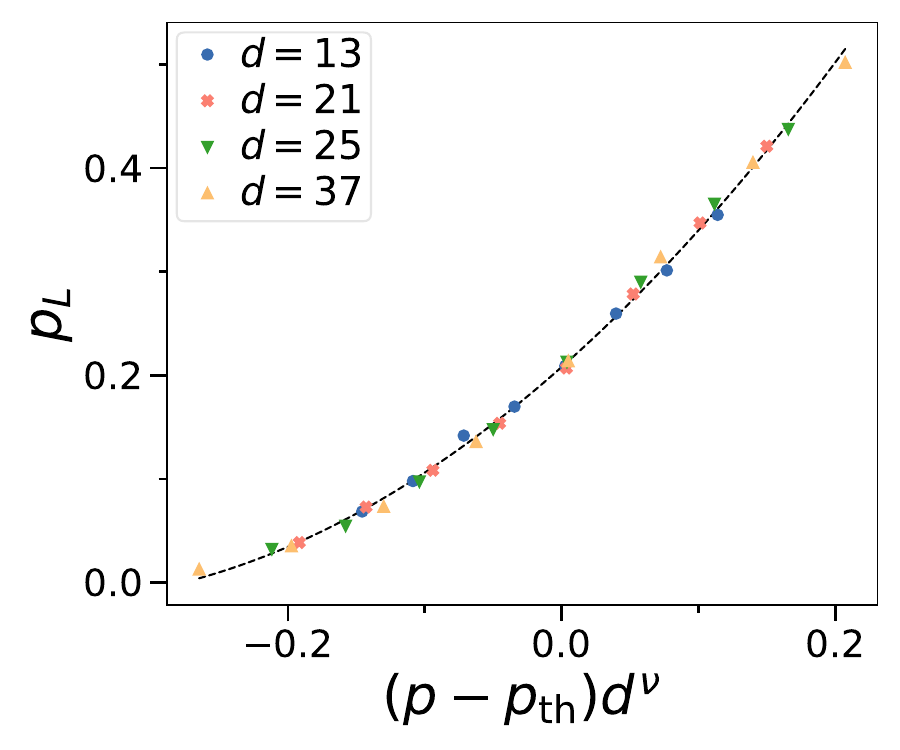}
	\caption{Threshold for Color code using our \idg{a,b} MaxSAT solver and \idg{c,d} BP-OSD. \idg{a,c} We plot logical error rate $p_\text{L}$ against physical error rate $p$ for depolarizing noise for different code distances $d$. We find threshold $p_\text{th}=15.23\pm0.05\%$ for MaxSAT and $p_\text{th}=13.36\pm0.06\%$ for BP-OSD. 
 \idg{b,d} Rescaled error plot around $p_\text{th}$  with exponent $\nu=0.68$ for MaxSAT and $\nu=0.57$ BP-OSD, where the dashed line is the fitted critical scaling ansatz.
	}
	\label{fig:ColorLegacy}
\end{figure*}

We show the threshold  for our MaxSAT decoder and BP-OSD for the Toric code  in Fig.~\ref{fig:Toric}. 
We also show also data on more values of distance $d$ on the Color code in Fig.~\ref{fig:ColorLegacy}.

\end{document}